\documentclass[12pt,a4paper,final]{iopart}
\usepackage[utf8]{inputenc}

\usepackage{iopams}
\usepackage{graphicx}
\usepackage[breaklinks=true,colorlinks=true,linkcolor=blue,urlcolor=blue,citecolor=blue]{hyperref}
\usepackage{subcaption}
\usepackage{cite} %to combine citations in output as of e.g. [1-4]
\begin{document}

\title[Reproducibility of ‘COST Reference Microplasma Jets’]{Reproducibility of ‘COST Reference Microplasma Jets’}

\author{F Riedel\textsuperscript{1}, J Golda\textsuperscript{2,3}, J Held\textsuperscript{3}, H Davies\textsuperscript{1,4}, M W van der Woude\textsuperscript{4}, J Bredin\textsuperscript{1}, K Niemi\textsuperscript{1}, T Gans\textsuperscript{1}, V Schulz-von der Gathen\textsuperscript{3}, D O’Connell\textsuperscript{1}}
\address{\textsuperscript{1}York Plasma Institute, Department of Physics, University of York, York YO10 5DD, UK}
\address{\textsuperscript{2}Institute of Experimental and Applied Physics, Kiel University, 24098 Kiel, Germany.}
\address{\textsuperscript{3}Experimental Physics II, Ruhr-Universität Bochum, 44801 Bochum, Germany}
\address{\textsuperscript{4}York Biomedical Research Institute, Hull York Medical School, University of York, York YO10 5DD, UK}
%\ead{deborah.oconnell@york.ac.uk}

\begin{abstract}
Atmospheric pressure plasmas have been ground-breaking for plasma science and technologies, due to their significant application potential in many fields, including medicinal, biological, and environmental applications. This is predominantly due to their efficient production and delivery of chemically reactive species under ambient conditions. One of the challenges in progressing the field is comparing plasma sources and results across the community and the literature. To address this a reference plasma source was established during the ‘Biomedical Applications of Atmospheric Pressure Plasmas’ EU COST Action MP1101. It is crucial that reference sources are reproducible. Here, we present the reproducibility and variance across multiple sources through examining various characteristics, including: absolute atomic oxygen densities, absolute ozone densities, electrical characteristics, optical emission spectroscopy, temperature measurements, and bactericidal activity. The measurements demonstrate that the tested COST jets are reproducible, within the standard deviation, for each measurement.
\end{abstract}

\maketitle

\section{Introduction}
Atmospheric Pressure Plasma Jets (APPJs) have attracted significant interest due to their application potential, such as in potential cancer treatments and wound healing \cite{Hirst2015,Packer2019,Privat-Maldonado2018a,Privat-Maldonado2018b,Metelmann2015,Vandamme2012,Canal2017,Privat-Maldonado2019a,Privat-Maldonado2019b,Privat-Maldonado2016,Laroussi2002,Hirst2016,VanGils2013,Weltmann2008,Alkawareek2012}, plasma chemical \cite{Xu2020,Gorbanev2017,Gorbanev2016} and material synthesis \cite{Barwe2015,Mariotti2010}, and surface modifications like thin film deposition \cite{Benedikt2007,ReuterRR2012}, etching of photoresist \cite{West2016a}, and pre-treatment of plastic surfaces \cite{Shaw2016}. The main motivation of these plasma sources for technological applications stems from their ability to generate and deliver reactive atomic and molecular species (both long- and short-lived), along with other active components such as UV, charged particles, and electric fields, under ambient conditions to a target. Low temperature plasmas can stimulate specific biological responses, this is not only, but at least significantly due to the fact that low temperature plasma generated Reactive Oxygen Nitrogen Species (RONS) are the same as the RONS produced endogenously in the human body \cite{Graves2012,Graves2014a}. These mediate many physiological processes, such as cell-cell signaling, immune response, wound healing and cell death processes, so therefore the plasma produced species are expected to mimic the functions of their native counterparts. Some of these key reactive species include atomic oxygen and nitrogen, hydroxyl radicals, nitric oxide, singlet oxygen, ozone and hydrogen peroxide. The role of other plasma components, for example electric fields and UV, and more importantly the synergies between these components are also recognised as key to exploit in order to develop effective therapeutics. In fact, both RONS and electric fields individually are already known to play vital roles in existing therapeutics \cite{Campana2019,Yang2018,Chung2020}, and the ability of plasmas to directly generate and simultaneously deliver these offers significant advantages and potential. In this context, it is crucial to elucidate mechanisms, and to that end efficient and accurate transfer of knowledge across the community is critical in order to accelerate the pace of research. 

Due to the diverse application potential, the rapidly developing field, and technological need many atmospheric pressure plasma sources have been developed world-wide for application and fundamental science. These have proven to be efficient and successful for various means, however, what remains a significant challenge is the comparison across plasma sources. This results in inefficient scientific progress, as each research team either needs to conduct lengthy characterisation of each source, or has limited access to complex, expensive diagnostic and/or simulation capabilities to do so. This can, therefore, result in a lot of redundancy of research, but more critically, without correlation of plasma parameters, causation mechanisms of plasma-induced biological processes is extremely difficult - if not impossible. 

To help overcome these difficulties, within the European Cooperation for Science and Technology (COST) Action MP1101 ‘Biomedical Applications of Atmospheric Pressure Plasmas’ \cite{COSTaction}, a reference plasma source for atmospheric pressure plasmas, the COST Reference Microplasma Jet, or the so-called ‘COST jet' was introduced \cite{Golda2016}. The aim of such a source is to have a well-characterised plasma where the literature and knowledge from various teams can assist with elucidating plasma and application mechanisms. Until this little effort was made to establish a reference atmospheric pressure plasma jet with a freely available design \cite{COSTjetDrawings}, as was conducted for low pressure plasmas with the Gaseous Electronics Conference (GEC) RF Reference Cell \cite{Olthoff1995,Hargis1994}

The aim of the study presented here is to establish the variability between COST plasma jets, from the same manufacturing batch, and the reproducibility of each individual source. The motivation for this is two-fold: a reference source is only as good as how comparable it is to other sources, and how reproducible it is to itself. This should also serve as an aid for plasma source developers to help better understand the origin of variability in various plasma parameters.

\section{Background}
The COST jet is driven with a radio frequency waveform, capacitively coupled with parallel stainless steel electrodes, sandwiched between glass panes to confine the gas flow between the electrodes. It is typically operated with a noble gas flow e.g. helium and small admixtures of molecular gases such as oxygen, nitrogen or water vapour. The geometry has been designed to provide good optical diagnostic access to the plasma core \cite{Wijaikhum2017,Knake2010a,Wagenaars2012,Maletic2012} as well as the jet/effluent region, and its geometry is also well suited for simulation and modelling of the plasma \cite{Murakami2014,Murakami2013,Murakami2013a,Niemi2010,Waskoenig2010b}. The 30 mm plasma channel ensures, that for typical operating gas flows in the order of slm, the various chemical species have evolved to steady state well before the nozzle \cite{Knake2010a,Hemke2011,Schroeter2018}. 

The cross-field geometry configuration of the plasma is such that the plasma jet, or effluent, consists of neutral species, and UV radiation can also escape and be transported to the target \cite{Golda2020}. Since the electric field between the electrodes is perpendicular to the gas flow, the charged particles and electric field rapidly decay outside the core plasma, therefore leaving the jet region devoid of charged particles and electric fields \cite{Walsh2008}. This means that the resultant plasma jet, or effluent, are not susceptible to typical surrounding ambient electrostatics and electrodynamics, while in comparison plasma jets containing charged particles and exhibiting relatively high electric fields, can be very susceptible to surrounding ground and target type \cite{Sobota2014,Lietz2018,Klarenaar2018,Darny2017,Kovacevic2018}.

There has been a significant body of research carried out on the COST jet, and its predecessors, including diagnostics, simulations \cite{Kawamura2014} and modelling \cite{Lazzaroni2012,Lazzaroni2012a,Niemi2013a}, on the electron dynamics, plasma sustainment mechanisms \cite{Shi2005,Iza2007,Liu2008,Schaper2011, Schulz-vonderGathen2008,Duennbier2015,Sakiyama2007,Hemke2013,Bischoff2018,Gans2010}, and chemical kinetics \cite{Schroter2018,Schroeter2018}, including reactive species production in the bulk plasma and jet regions \cite{Kelly2014,Willems2019}. The plasma has been applied for various applications and additionally advanced tailoring concepts \cite{Waskoenig2010a,ONeill2012,Kwon2013,Gibson2019, Liu2009,McKay2010,Dedrick2017,Kelly2014a,Korolov2019} are been developed and employed for improved control over the reactive species generation and treatments. Improvements on efficient power delivery and electrical characterisation have also been performed \cite{Marinov2014,Beijer2016,Golda2019}. These detailed characterisations and investigations can help inform future research on the COST jets and also adapted more generally for other atmospheric pressure plasma sources.

In order to control the production of reactive species within the plasma, molecules are purposely admixed to the feed gas. This provides improved control over reactive species generation, compared with relying on ambient oxygen and nitrogen molecules \cite{Schroeter2018}. Knake et al. \cite{Knake2008,Knake2008b} reported that the atomic oxygen production is most efficient with an admixture of 0.5\,\% to 0.6\,\%  molecular oxygen. This motivates the molecular admixtures used in the presented work. Various reactive species have been measured and simulated using different techniques including atomic oxygen \cite{Knake2010,Maletic2012,Niemi2013,Niemi2009,Greb2014}, hydroxyl \cite{Schroter2018,Benedikt2016}, singlet delta oxygen \cite{Sousa2011}, ozone \cite{Wijaikhum2017}, hydrogen peroxide \cite{Vasko2014}, atomic nitrogen \cite{Wagenaars2012,Niemi2013}, nitric oxide \cite{Douat2016}, helium metastables \cite{Niemi2011, Spiekermeier2015,Niermann2011}. These reactive species can propagate varying distances beyond the plasma nozzle, with some surviving up to several centimetres \cite{Kelly2015}. The role of synergies between UV and reactive species, compared with their individual influence, has been investigated using an extended X-jet configuration and identified as important \cite{Schneider2011,Schneider2011a,Lackmann2013,Edengeiser2015,Golda2020}. In general for plasma treatments, the flux of both these components is important to consider, as is the heat and gas dynamics impacting on surfaces \cite{Kelly2015}.

Gorbanev et al. determined the origin of species in plasma treated liquids and found that most reactive species detected in the liquid phase originated in the plasma gas phase and were subsequently transported into the liquid \cite{Gorbanev2018}. While Hefny et al. highlight the important role of solvated O atoms in acqueous solutions \cite{Hefny2016}. Investigations have also started to elucidate mechanisms in physiological solutions \cite{Lackmann2018,Lackmann2019,Jirasek2019,Gorbanev2019}. Treatment of different biological systems have been conducted for efficacy purposes, but also to clarify mechanisms of plasma action. Treatments on different cancer cell types \cite{Gibson2014,Vermeylen2016,Bekeschus2017,Privat-Maldonado2018b}, different DNA origami nanostructures \cite{OConnell2011,Miller2020,Edengeiser2015}, and antibacterial action, including resistance mechanisms \cite{Krewing2019} have been investigated. The COST jet has also been applied for various other applications \cite{Gorbanev2019a}, including the development of new chemical and material synthesis and processing such as epoxidation \cite{Xu2020}, etching \cite{West2016a}, thin film deposition \cite{ReuterRR2012}, and surface modifications \cite{Shaw2016}. 

The aim of the study presented here is to investigate the repeatability across different COST reference microplasma jets. Therefore the plasmas, produced by four devices, are compared for different parameters. These include power characteristics, gas and substrate surface temperature, optical emission spectroscopy, ozone densities, atomic oxygen densities, and bactericidal  activity.

\section{Setup and Diagnostics}

\subsection{Plasma Source}

In this study four identically constructed and equipped exemplars of the ''COST reference microplasma jet", as specified in \cite{Golda2016}, were investigated. Each device consists of (a) the head, which includes the stainless steel electrode assembly between two quartz glass windows forming a plasma channel of 30\,mm length and 1\,mm\,x\,1\,mm cross section, and (b) the housing, which comprises the LC resonance based radio-frequency power coupling circuit \cite{Marinov2014}, a capacitively coupled voltage probe, and a resistive current probe.

A commercial 13.56\,MHz radio-frequency generator and external manual matching network unit (Coaxial Power Systems, RFG50 and MMN150) were used to operate the COST jet devices, with a 50\,$\Omega$ BNC coaxial cable of 0.5\,m length between matching unit and jet. The length of the feed gas tube that is exposed to moist ambient air whenever the jet is not in operation was kept as short as possible, in order to decreases the time for the jet to reach steady-state operation. The feed gas for all later experiments was chosen as 1\,slm helium flux with 0.5\,\% oxygen admixture (purity grade N4.6 for helium and N5.0 for oxygen). This ensures a high flux of generated reactive species to the sample, while keeping evaporation of wet biological samples tolerable \cite{Knake2008b, Wagenaars2012}.

In order to ensure a valid comparison, all four COST reference microplasma jets were initially cleaned with standard solvent in an ultrasonic bath for the presented investigation, because each of them had an unknown prior history, e.g. time and conditions of operation, at different universities and institutes. Each jet was operated with the exact same equipment: radio-frequency generator, matching network, coaxial cables and connectors, and gas mass flow controllers. The same applies for all measuring equipment, e.g. digital oscilloscope, spectrometer, ozone monitor, lasers, external electrical probes. Prior to any experiment, the gas lines were flushed for 30\,min, then the plasma jet was ignited and run for a 30\,min warm-up duration. After any change of the operational parameters, externally applied rf power or gas flow, a stabilisation time of 10\,min was observed before conducting the next measurement. The laboratory conditions were controlled to within $22\pm0.5\,^\circ\rm{C}$ room temperature and 50$\pm$10\,\% relative humidity for all measurements.

\subsection{Measurements}

\paragraph{Dissipated Electrical Power}

The COST-Jet incorporates miniaturised electrical probes inside its housing, i.e. a capacitively coupled voltage probe and a resistive current probe, which allow a precise measurement of the actual electrical power dissipated in the plasma \cite{Golda2019}. The output of both probes was simultaneously recorded by an oscilloscope (Agilent Technologies, Infiniivision DSO-X 2004A, 8\,bit, 2\,GSps, 70\,MHz) using 50\,$\Omega$ coaxial cables (Thorlabs, CA2612) of equal type and length, as an average over 1024 consecutive recordings. The data was sent to a computer where the deposited power was calculated according to 

\begin{equation}
    P_{plasma}=U_{rms}*I_{rms}*\cos(\phi-\phi_{ref}),
\end{equation}

with $U_{rms}$ and $I_{rms}$ the effective values of voltage and current, $\phi$ the phase shift between voltage and current, and $\phi_{ref}$ the instrumental reference phase shift determined from measurements without plasma. Since the internal probes are located directly at the electrodes, their readings do not need to be corrected for parasitic power consumption occurring elsewhere in the circuit (not in the plasma), as e.g. in \cite{Duennbier2015,Hofmann2011}. A detailed error analysis of the power measurement method can be found in \cite{Golda2019}.
The internal voltage probe of each COST jet needs to be calibrated \cite{Golda2016}, e.g. here by using a commercial external voltage probe (Tektronix, P5100A).

\paragraph{Effluent Gas Temperature}

The gas temperature of the jet's effluent as a crucial parameter for the treatment of biomedical or heat sensitive samples was measured with a K-type thermocouple that was placed 3\,mm in front of the jet's nozzle. No evidence was found that the thermocouple measurement was influenced by the radio-frequency electro-magnetic radiation from the electrode assembly. Each temperature measurement as an average over 5\,min, taken alongside the electrical power measurement, results in a mean value and a standard deviation, which reflects changes like airflow and room temperature inside the lab.

\paragraph{Surface Temperature}

The spatially resolved surface temperature on an artificial sample placed at various distances from the jet’s nozzle was measured with a thermal camera (Agilent, U5855A). The chosen sample is a standard microscope slide out of chemically inert quartz glass. The sample’s front surface was roughened by shot blasting to minimise direct reflections from other heat sources. The thermal emission coefficient of this surface was measured with respect to the known emission coefficient of a black infrared sticker.

\paragraph{Optical Plasma Emission}

The optical emission from the centre of the plasma channel is measured with a fibre coupled spectrometer (Ocean Optics, HR4000CG and QP600-2-SR-BX) using the reference fibre spacer of the COST jet as alignment tool for positioning the fibre tip, see \cite{Golda2016}. The fixed-configuration spectrometer covers the spectral range from 200\,nm to 1100\,nm with a spectral resolution of about 0.5\,nm. We focus on measuring intensity ratios of the dominant atomic emission lines, He(706\,nm), O(777\,nm), and O(844\,nm), for comparing the four different COST jets.

\paragraph{Ozone Densities}

The ozone density in the far-effluent of the COST jet was measured with a commercial ozone gas detection monitor (2B Technologies, Model 106-L) based on 254\,nm UV absorption. The gas output of the jet was sucked into the ozone monitor via a glass funnel close to the jet’s nozzle and through a 1\,m long PFA plastic tube by the internal pump of the monitor at nominal flow rate of 1\,slm.

\paragraph{Atomic Oxygen Density}

Atomic oxygen ground state densities in the near effluent of the COST jet, here at 1\,mm distance from the nozzle and when operated with standard feed gas of 1\,slm helium with 0.5\,\% oxygen admixture, were measured by means of two-photon absorption laser induced fluorescence (TALIF). An absolute calibration was carried using the technique detailed in Niemi et al. \cite{Niemi2001, Niemi2005}. In this case, an evacuated gas cell was filled with a defined pressure of xenon, with a similar TALIF scheme to atomic oxygen. The challenging aspect of this established method is determination of the effective lifetime of the laser excited atomic oxygen state. This lifetime is typically in the order of picoseconds to a few nanoseconds due to significant and inhomogeneous collisional quenching within the effluent of the COST jet penetrating into ambient air.

Two different spectrally widely tunable OPO/OPA laser systems with inbuilt frequency doubling and mixing stages were used in this study. For all presented TALIF measurements, the fluorescence is detected perpendicular to the laser beam, but in the same horizontal plane, while the COST jets were mounted upright, with the effluent directed towards a fume extraction hood. 

A pico-second laser system (Ekspla, PL2251B, APL2100, and PG411), able to provide up to $300\,\mu\rm{J}$ energy within 30\,ps pulse duration with a repetition rate of 10\,Hz at the required wavelengths around 225\,nm, was used as an excitation for the TALIF schemes. The UV output beam was intentionally attenuated and focused with an $f=30\,\rm{cm}$ silica lens about 3\,cm behind the COST jet effluent, to keep the local power density below the onset of various signal saturation effects. The fluorescence volume was imaged about 1:1 by a doublet of achromatic lenses (1\,inch diameter, $f_{tot} = 40$\,mm) through interference filters of 10\,nm bandwidth onto the chip of an intensified charge-coupled device camera (Stanford Computer Optics, 4-Picos with S25IR photo-cathode and $780 \times 580$\,pixels of $8.3\,\mu$m square size). This laser system in combination with the camera’s minimal gate width of 200\,ps was used to measure the exponential decay of the fluorescence radiation from the laser excited states in the COST jet effluent and in the reference gas cell under low pressure, respectively. Our measurements result in an effective O($3p\,^3P_J$) state lifetime of $4.24\pm0.07$\,ns, which is about 8\,times shorter than the corresponding natural lifetime of 35.1\,ns.

A second TALIF setup includes a more conventional nanosecond OPO/OPA laser system (Continuum, Surelite EX and Horizon OPO) providing 225\,nm pulses of up to 5\,mJ energy in about 4\,ns duration at the same repetition rate of 10\,Hz, and a different ICCD camera (Andor, IStar with -73 photocathode and $1024 \times 1024$\,pixels of $13\,\mu$m square size) with a longer minimal gate width of 2\,ns, in an otherwise similar detection setup. This second setup, since it was offering a lower pulse-to-pulse laser energy fluctuation than the first, was used to measure the temporally and spectrally integrated TALIF signals, for atomic oxygen in the effluent and for xenon in the reference gas cell, from which the stated atomic oxygen density values were derived.

%\subsection{Calibration factor deviation}
%The determination of the calibration factor is crucial part for the COST jets because of its capacitively coupled voltage probe. In the following analysis only statistical errors that show the differences between the jets will be shown. The calibration factors were obtained at the Ruhr-University Bochum, Germany, and used throughout this study. Table \ref{tab:Calibration-factors} shows the calibration factor for each jet. The calibration was done by attaching a voltage probe to the driven electrode and comparing this voltage to the voltage obtained from the internal voltage probe 30 times. The voltage signal is averaged 2048 times which leads to a total average of 61440 times. Thus leaving only a systematic errors for the voltage measurements. It is crucial that the case is closed during the calibration.

%\begin{table}[!h]
%\begin{centering}
%\begin{tabular}{|c|c|c|c|c|}
%\hline
%Jet & Calibration factor  & from \cite{Golda2016}& York %calibration&deviation (\%)\tabularnewline
%\hline
%\hline
%A & 2193$\pm$50 &  & 2308 &5 \tabularnewline
%\hline
%D & 2013$\pm$50 &  & 2201&8.5\tabularnewline
%\hline
%E & 2916$\pm$50 & 2630$\pm$50 &3127&6.7 \tabularnewline
%\hline
%G & 2728$\pm$50 & &2898&5.9  \tabularnewline
%\hline
%\end{tabular}
%\par\end{centering}

%\protect\caption{Calibration factors for each jet.\label{tab:Calibration-factors}}
%\end{table}

\paragraph{Bactericidal Assay}

The bactericidal assay was carried out as described previously in \cite{Privat-Maldonado2016}. Briefly, single colonies of non-pathogenic \textit{Escherichia coli} K-12, MG1655, were cultured in Luria-Bertani broth (LB, 10\,g/L), until the late logarithmic growth phase. The optical density (OD) of the bacteria was then adjusted to $OD = 0.02$ at 600\,nm. Approximately $8 \times 10^5$ colony forming units (cfu) were transferred to LB agar petri dishes (LB, supplemented with 17.5\,g/L agarose), and spread using glass beads to ensure even coverage of the plates with \textit{E. coli}. The bacterial plates were allowed to dry.

For plasma treatment, the COST jets were operated in downwards orientation inside a reasonably large Perspex plastic box with constant fume extraction at the top. The box shields the experiment from changing air movements within the lab, while the fume extraction prevents a build-up of reactive species inside the box. Before the 2\,min long plasma treatment, the lid of the petri dish was removed, plate was placed so that the top of the agar was at a distance 5\,mm below the plasma jet nozzle. Afterwards, the plate was immediately removed, the lid put back on to prevent contamination, and the plate returned into the incubator (at $37\,^\circ\rm{C}$) for overnight. On the following day, the plates were imaged using a scanner (Epson, Perfection V750 Pro) to measure the area of inhibition (AOI) and to count the number of survivor colonies using the \textit{ImageJ} software \cite{ImageJ}.

Care was taken to follow the exact same experimental protocol: Autoclaved LB agar was always allowed to cool-down to $55\,^\circ\rm{C}$ in a water bath, then 20\,mL was poured into each petri dish, and the plates were allowed to dry in a laminar flow hood for the same duration. Also, all bacterial cultures were derived from single colonies, cultured over same duration and under same conditions, and the bacterial concentration was kept constant for all experiments.

\section{Results}

\subsection{Power Characteristics}

Figure \ref{fig:power-voltage} shows a measured characteristic of dissipated electrical power versus effective voltage as obtained from a subsequent forward/backward sweep of the external generator power. The black dots are those measured from one of the COST jets. The blue and red areas for forward and backward sweep, respectively, and the overlap in purple, indicate the standard deviation found among the four COST jets. Hysteresis effects are visible, e.g. the plasma ignites into a stable homogeneous operational mode at about 190\,$V_{rms}$, which can be sustained down to about 160\,$V_{rms}$ after ignition. At a voltage of about 355\,$V_{rms}$, the jump to higher power and lower voltage marks the spontaneous transition into an operational mode with one constricted filament between the electrodes, which is clearly distinguishable by naked eye observation. This mode is not desired as the filament is unstable and the power is constricted to hot-spots on the electrode surfaces leading to damage as well as excessive gas heating on short time scales. For extinguishing the filament, the voltage/power needs to be reduced well below the onset of this mode transition.
The four COST jets showed a stable homogeneous operation from 180 to 350\,$V_{rms}$ and corresponding plasma power from 0.18\,W to 2.5\,W. The relative standard deviation among the four COST jets in terms of plasma power increase with effective voltage, but stays below 15\,\% as indicated by the blue and purple overlap area in figure \ref{fig:power-voltage}, when the constricted mode is avoided.         
A detailed error analysis of electrical power measurements on COST jet devices has been presented in \cite{Golda2019}. Our investigation falls into the second of the three different scenarios that were considered in this reference, i.e. comparing different devices but using the same rf power equipment, resulting in a predicted overall relative error of about 10\,\%.  
The observed deviation of 15\,\% between the investigated four COST jet devices is slightly larger than the predicted uncertainty of 10\,\% for the electrical power measurement. The cause most likely reflects the result of small manufacturing tolerances for the electrode gap distance and alignment (within $\pm0.1\,\rm{mm}$) and the actual electrode surface conditions prone to physical roughness and surface coverage (humidity and oxides), which depend on previous operational conditions (feed gases), or even possible damage from operation in filamentary mode.

\subsection{Effluent gas temperature} \label{effluenttemp}

\begin{figure}[ht!]
\centering
\begin{subfigure}{0.49\textwidth}
\includegraphics[width=1.0\linewidth]{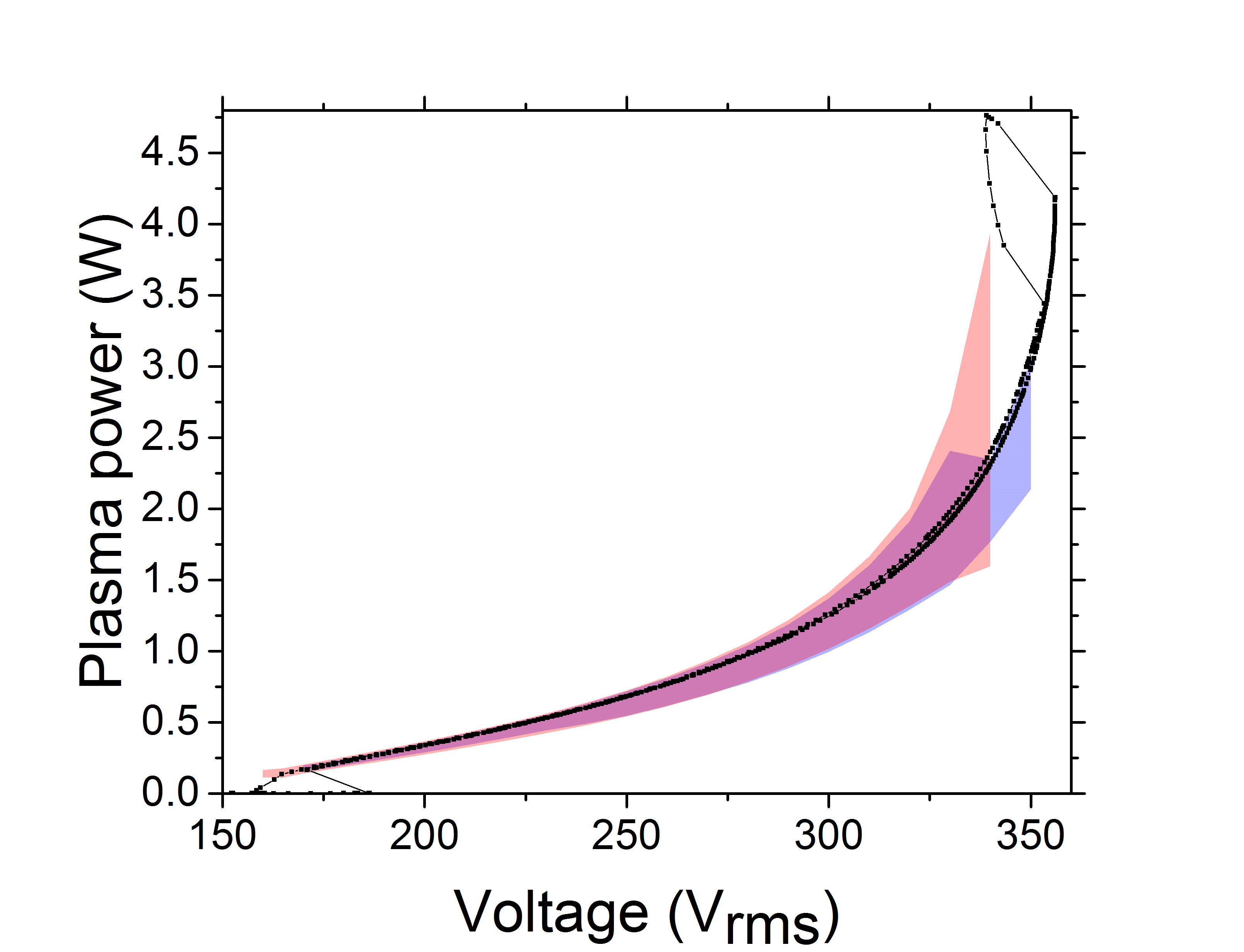}
\caption{}
\label{fig:power-voltage}
\end{subfigure}
\begin{subfigure}{0.49\textwidth}
\includegraphics[width=1.0\linewidth]{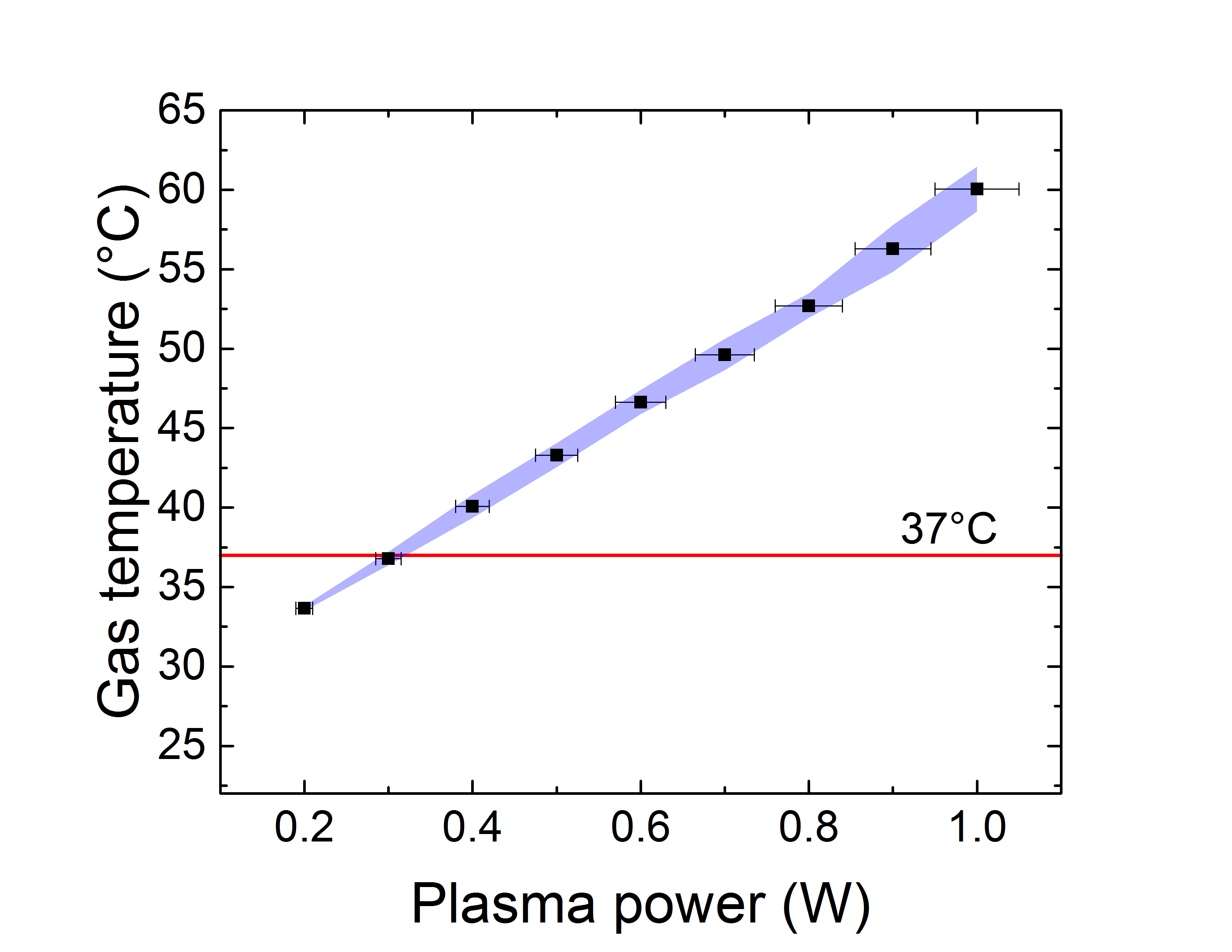}
\caption{}
\label{fig:tempvspower}
\end{subfigure}
\caption{(a) Measured characteristics of plasma power versus effective voltage, and (b) measured effluent gas temperature at 3\,mm distance from the nozzle versus plasma power, both for standard feed gas of 1\,slm He with 0.5\,\% oxygen admixture. Black dots indicate the result obtained from one of the COST jets. The coloured areas are the standard deviation between the four COST jets, explanation provided in text. The red horizontal line in (b) indicates the maximum permissible temperature for regular treatment of biological samples.}
\end{figure}

Figure \ref{fig:tempvspower} shows the measured gas temperature of the free flowing effluent at 3\,mm distance from the nozzle as a function of the plasma power for the jets investigated in this study. The blue area indicates the standard deviation among the four COST jets, staying below 3\,\% with respect to room temperature over the whole operational range. The corresponding error margins in y-axis stay within the indicated x-axis error margins for the derived plasma power.
The horizontal red line indicates the maximum permissible temperature of $37\,^\circ\rm{C}$ for regular treatment of biological samples, which implies that the plasma power should be kept below 0.3\,W for this particular distance.

\subsection{Surface Temperature} \label{surftemp}

\begin{figure}[ht!]
\centering
\includegraphics[width=0.9\linewidth]{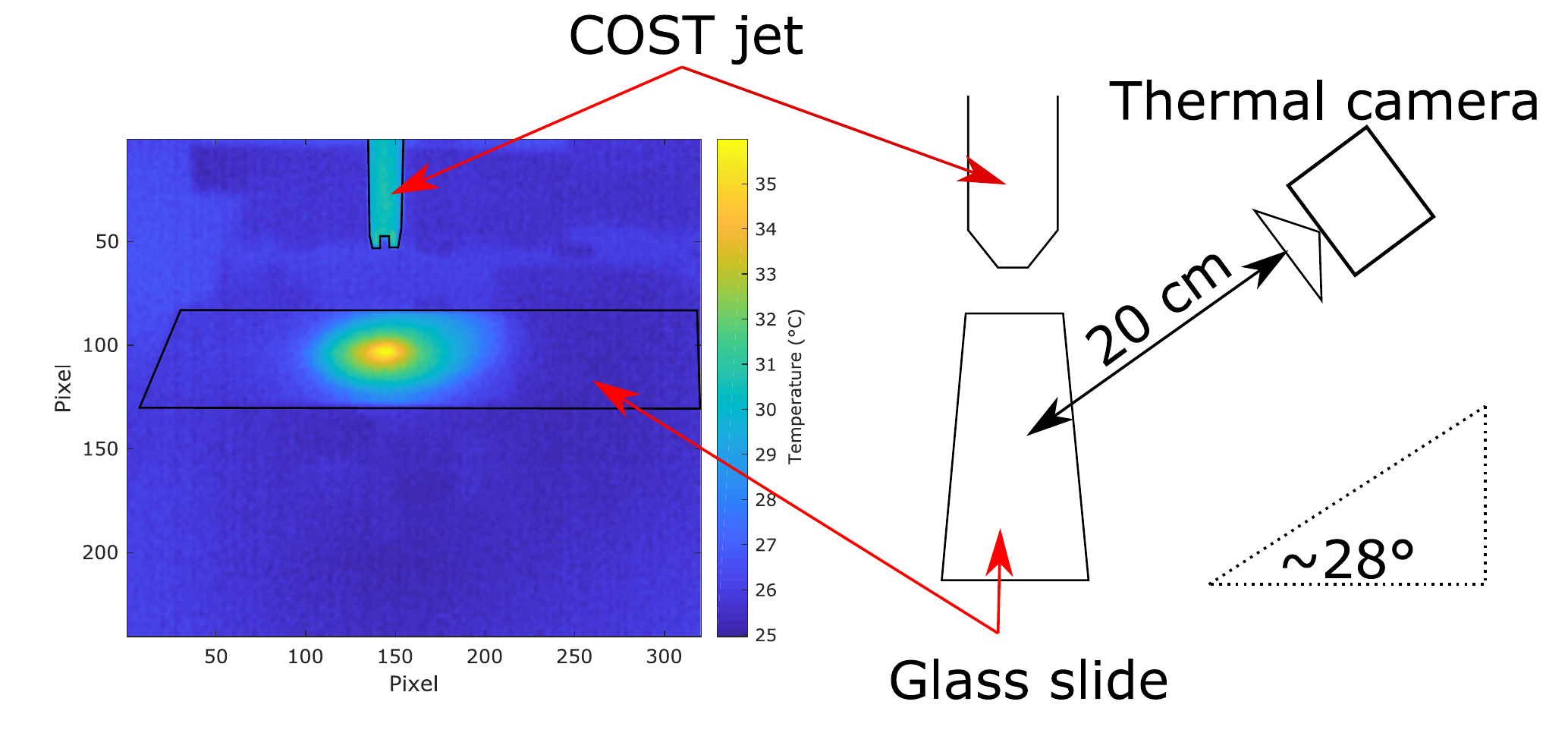}
\caption{(Left) Thermal image taken for standard feed gas (1\,slm\,He with 0.5\,\% oxygen admixture) and at reduced plasma power (0.3\,W). (Right) Schematic of the corresponding experimental setup.}
\label{fig:surfacetemp2}
\end{figure}

Figure \ref{fig:surfacetemp2} (Left) shows an example thermal image taken by the thermal camera, and (Right) the schematic of the measuring setup. The COST jet is mounted horizontally and pointing towards the microscope slide. The thermal camera is mounted on a tripod at a distance of about 20\,cm and at an angle of about $28\,^\circ$ with respect to the shot-blasted front surface of the microscope quartz slide. As expected, the thermal image shows that the spatial profile of the surface temperature exhibits a central maximum at the location where the jet axis hits the surface.

\begin{figure}[ht!]
\centering
\begin{subfigure}{0.49\textwidth}
\centering
\includegraphics[width=1.0\linewidth]{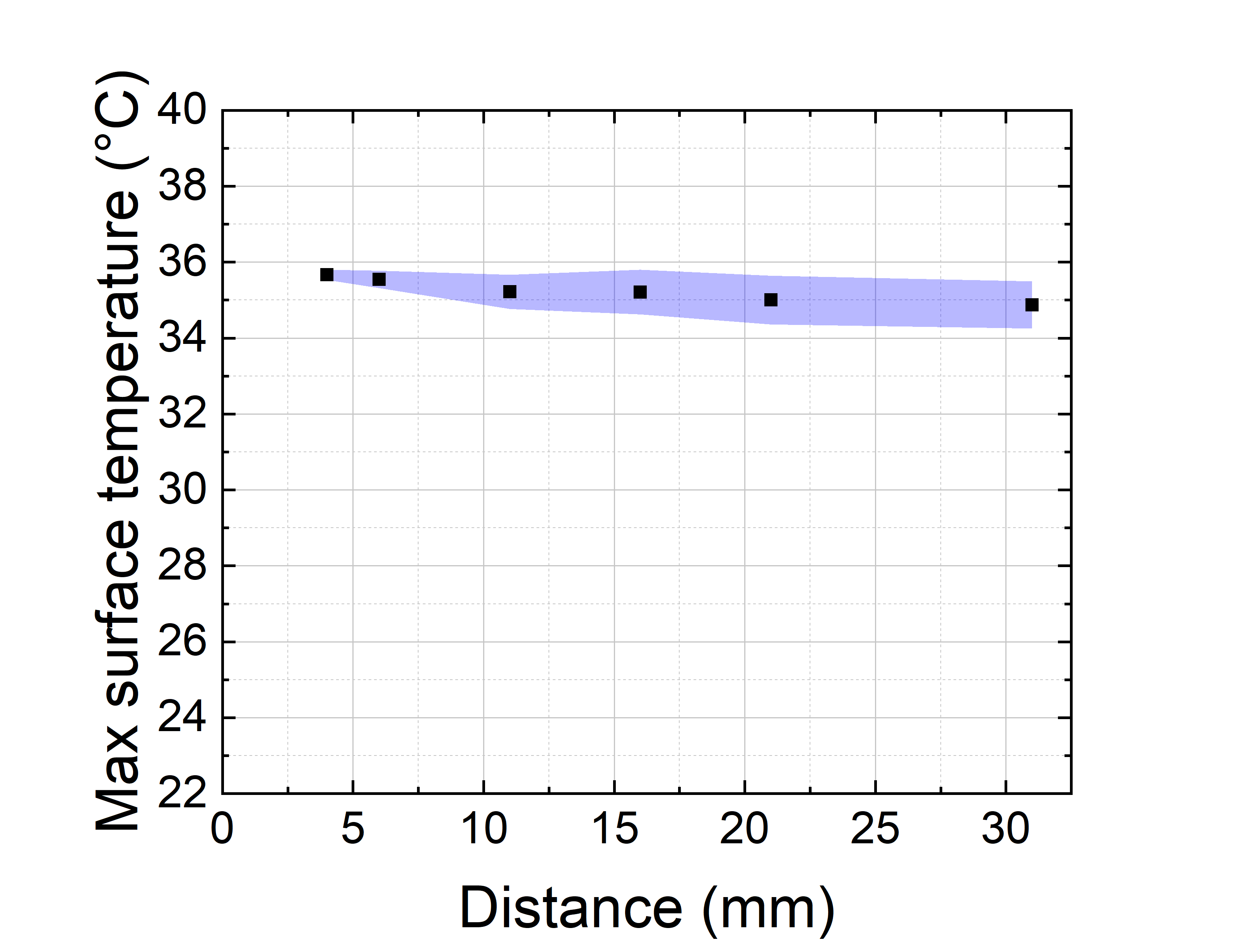}
\subcaption{}
\label{fig:surfacetemp}
\end{subfigure}
\begin{subfigure}{0.49\textwidth}
\centering
\includegraphics[width=1.0\linewidth]{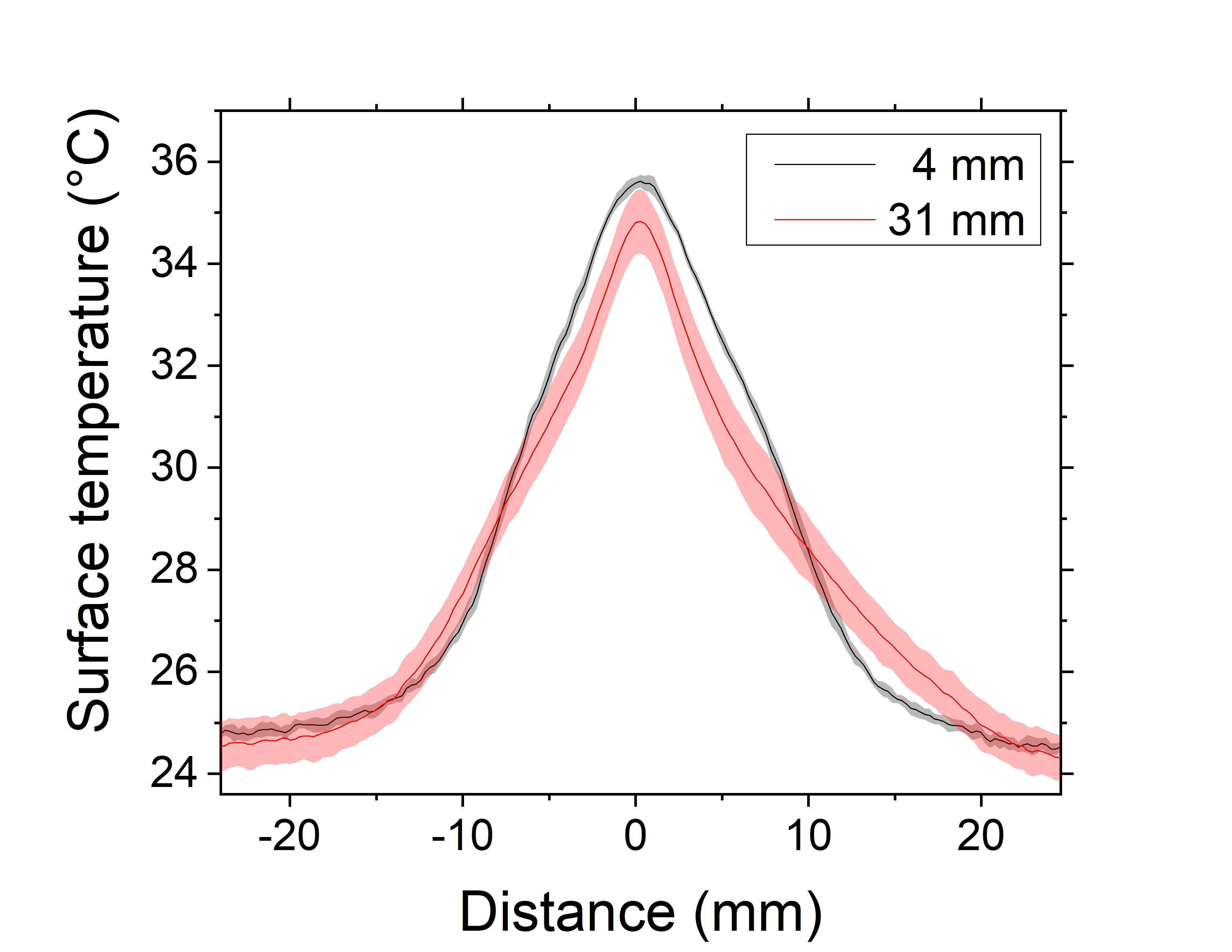}
\subcaption{}
\label{fig:heatprofilediffmaxdistances}
\end{subfigure}
\caption{(a) Measured maximum central surface temperature as a function of the distance between the jet nozzle and sample. (b) Measured lateral surface temperature profiles at different nozzle to sample distances. Both for standard feed gas (1\,slm\,He with 0.5\,\% oxygen admixture and at reduced plasma power (0.3\,W).}

\end{figure}

Figure \ref{fig:surfacetemp} shows the measured maximum central surface temperature as a function of the distance between jet nozzle and sample surface, for a plasma power of 0.300\,$\pm$\,0.015\,W. At each distance, three thermal images of the sample surface are taken for each COST jet, respectively. The black dots represent the mean values, and the blue area the standard deviation among the four COST jets. The measurement shows that the maximum surface temperature is decreasing only insignificantly, from about $36\,^\circ\rm{C}$ to about one degree less, over a distance of up to 30\,mm distance from the jet nozzle. The standard deviation of $1\,^\circ\rm{C}$ surface temperature is the uncertainty of the thermal imaging measurement, given air movements in the lab, from which we conclude that the four COST jets are fully comparable in this respect. In addition, the surface temperature and the gas temperature in the free flowing effluent measured at the same plasma power agree within the $1\,^\circ\rm{C}$ uncertainty.

Figure \ref{fig:heatprofilediffmaxdistances} shows the measured lateral surface temperature profiles. The black profile with grey margins represents the mean values and corresponding standard deviations among the four COST jets from a measurement at near distance of 4\,mm, while the red curve with rose margin indicate the corresponding quantities from a measurement at far distance of 31\,mm. The measurement shows that the spatial temperature distribution does not spread significantly with increasing distance up to 31\,mm distance, as expected from the measured marginal decrease of the on-axis surface temperature. The standard deviation among the four COST jets however increases with distance, because of the mentioned environmental influences. Our findings that the free flowing jet effluent in terms of temperature stays constricted over a distance of 30\,mm and produces a lateral sample surface temperature profile of about 16\,mm full-width at half-maximum (FWHM) are supported by the spatially resolved thermocouple measurements in \cite{Knake2008} and Schlieren imaging in \cite{Kelly2015}.

\subsection{Optical plasma emission}

Optical plasma emission is a very useful external diagnostics to monitor plasma stability and reproducibility. It depends on various internal plasma parameters, in particular species composition, electron density and electron energy distribution function. While it is challenging to quantitatively distinguish between details of the origin of changes, it is highly sensitive to overall changes.

\begin{figure}[ht!]
\centering
\includegraphics[width=0.539\linewidth]{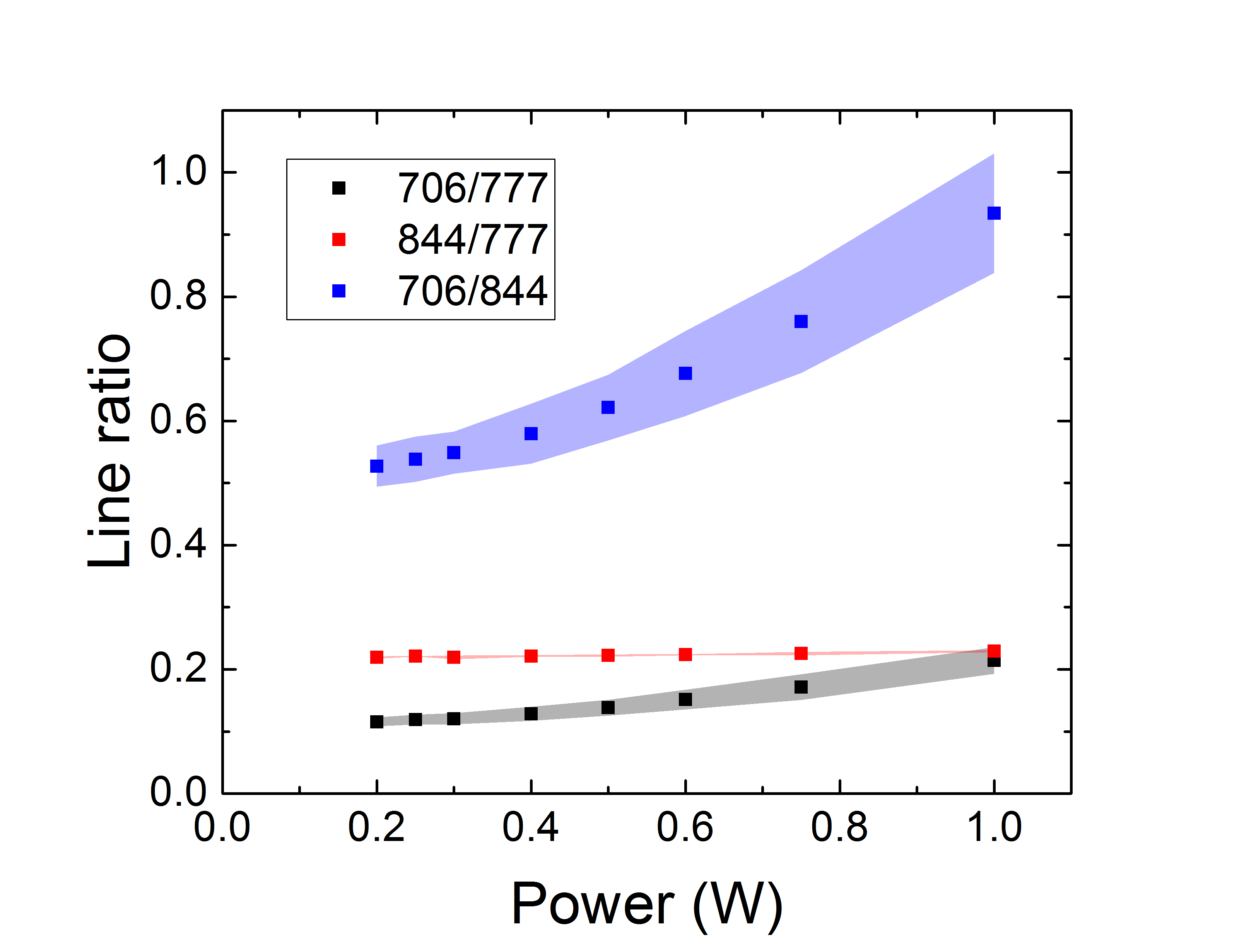}
\caption{Intensity ratios from optical emission measurements as a function of plasma power for standard feed gas of 1\,slm He with 0.5\,\% oxygen admixture. The considered atomic lines, He 706\,nm, O 777\,nm, and O\,844 nm, are labelled according to their wavelength.}
\label{fig:lineratios}
\end{figure}

The results of the optical emission spectroscopy (OES) measurements taken in the centre of the plasma channel are shown in figure~\ref{fig:lineratios} as intensity ratios for the most prominent atomic lines, He(706\,nm), O(777\,nm) and O(844\,nm), as a function of the dissipated plasma power. The solid squares represent the mean values and the shaded areas the standard deviations, respectively, found among the four COST jets.

The O(844\,nm)/O(777\,nm) intensity ratio stays almost constant over the investigated power range. Both intensity ratios, He(706\,nm)/O(777\,nm) and He(706\,nm)/O(844\,nm), show an increase with increasing plasma power. The standard deviations for the He(706\,nm)/O(777\,nm) and He(706\,nm)/O(844\,nm) line ratios are below 8\,\%. This indicates a very good agreement among the four COST jets, since this deviation is less significant than the uncertainty in the plasma power measurement. 

%, since both atomic oxygen lines exhibit similar threshold energies of about 11\,eV for direct electron-impact excitation from O atoms, and about 16\,eV for the lesser effective dissociative electron-impact excitation from O\textsubscript{2} molecules, see \cite{Niemi2009}, respectively. This ratio is therefore rather insensitive to the changes in atomic oxygen density and mean electron energy with increasing plasma power, when assuming no changes in the depopulation of the upper electronic states by spontaneous emission and collisional quenching. The found standard deviation of only 1\,\% among the four COST jets therefore represents the effective accuracy of our measurement.

%Both intensity ratios, He(706\,nm)/O(777\,nm) and He(706\,nm)/O(844\,nm), show an increase with increasing plasma power. From the previous statements, it is clear that both trends are similar. However, the fact that both ratios are increasing is not fully understood, since the atomic oxygen ground state density is increasing (as shown below), while the helium ground state density stays practically the same. This suggests (a) a significant increase in the fraction of hot electrons above the 22.7\,eV threshold energy for direct electron-impact excitation of the upper He($3s\,^3S_1$) state, although simulations predict a slight decrease of the mean electron energy with increasing plasma power \cite{Waskoenig2010b}, or (b) a major increase in the contribution of electron-impact excitation out of the metastable He ($2s\,^3S_1$) state, again unexpected, see \cite{Niemi2011}.

\subsection{Ozone density}

\begin{figure}[ht!]
\centering
\includegraphics[width=0.539\linewidth]{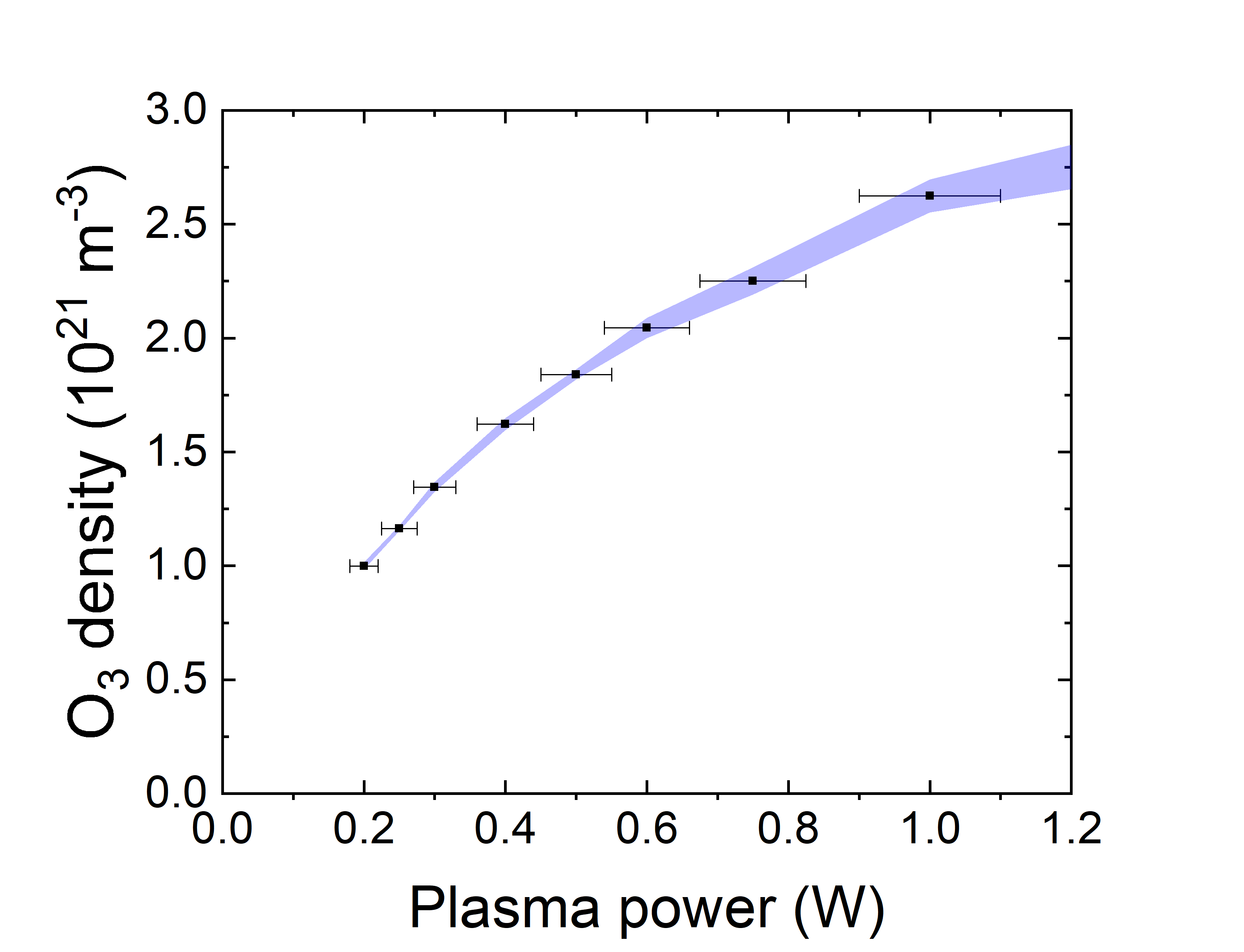}
\caption{Measured ozone density in the far effluent of the COST jets as a function of the plasma power for standard feed gas of 1\,slm He and 0.5\,\% oxygen admixture.}
\label{fig:ozonedens}
\end{figure}

Figure \ref{fig:ozonedens} displays the measured ozone density in the far effluent of the COST jet as a function of plasma power for a standard feed gas of 1\,slm helium and 5\,sccm oxygen. The black dots represent the mean values and the shaded area the standard deviation of the four COST jets. The ozone density increases under-linear with increasing plasma power from $1\times10^{21}\,\rm{m}^{-3}$ at 0.2\,W to $2.6\times10^{21}\,\rm{m}^{-3}$ at 1\,W. 

The formation of ozone in this type of plasma source has previously been investigated in detail using two-beam UV-LED absorption spectroscopy and numerical simulations \cite{Wijaikhum2017}. This revealed that the ozone density inside the bulk plasma source slightly decreases with increasing power. Here we use a simple ozone monitor in the far effluent where the plasma-produced atomic oxygen is already converted into additional ozone \cite{Ellerweg2012,Schulz-vonderGathen2007} through three-body recombination with molecular oxygen and helium \cite{Wijaikhum2017}. 

The observed increase in ozone density with increasing power in the far effluent can be explained by the increased production of atomic oxygen at elevated powers (see sub-section on atomic oxygen density below). The results for 0.5\,W are also in good agreement with previous measurements in the far effluent \cite{Ellerweg2012,Schulz-vonderGathen2007}. The ozone density is expected to be lower at shorter distances from the jet nozzle, due to less conversion of O and O\textsubscript{2} into ozone. The deviation of the ozone density between the jets stays below 3\,\% and is, therefore, less significant than the uncertainty of the plasma power measurement.

\subsection{Atomic oxygen density}

\begin{figure}[ht!]
\centering

\centering
\includegraphics[width=0.539\linewidth]{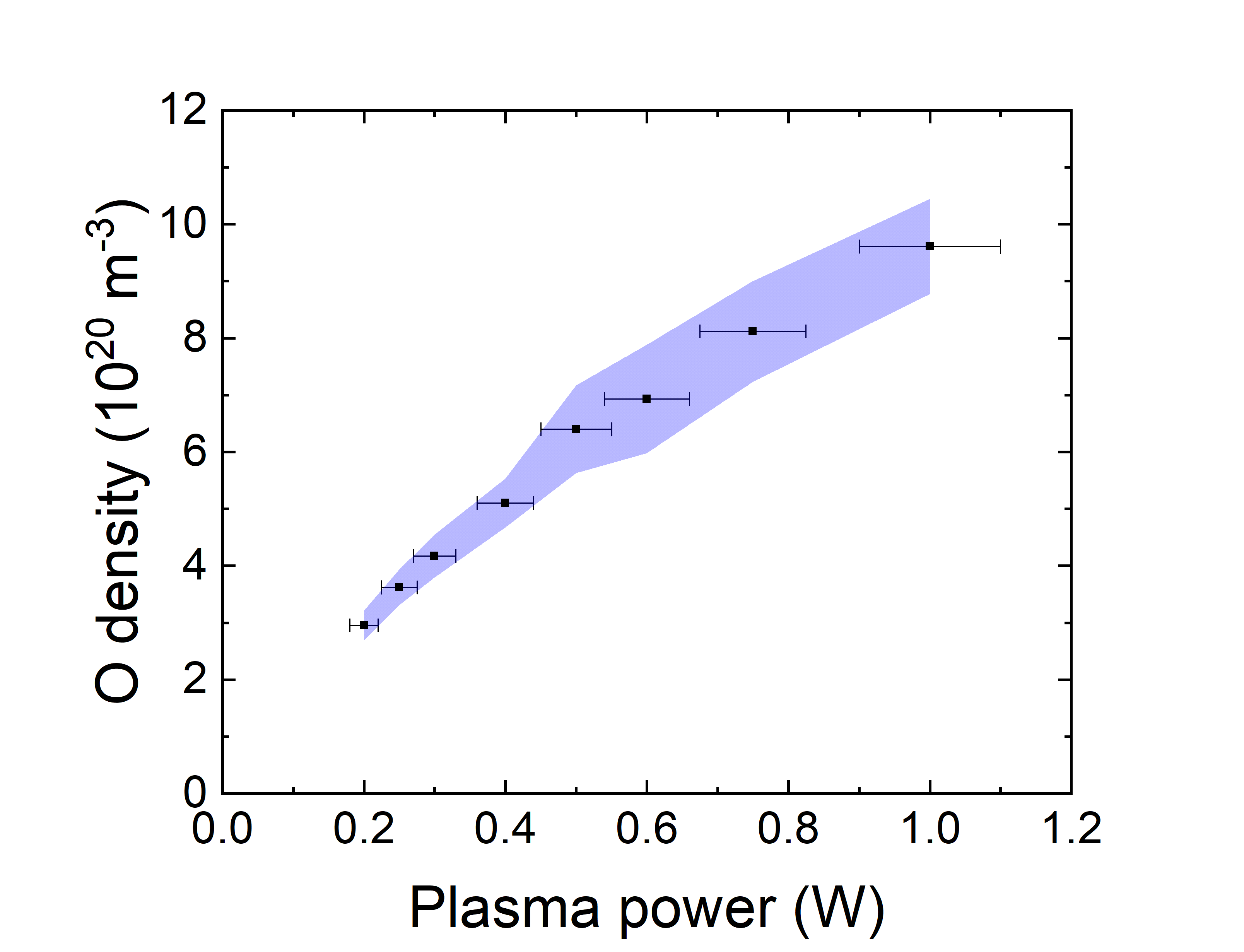}
\caption{Measured atomic oxygen density at 1\,mm distance from the COST jet nozzle as a function of the plasma powers for standard feed gas of 1 slm\,He with 0.5\,\% oxygen admixture.}
\label{fig:TALIFerrorbars}
\end{figure}

Figure \ref{fig:TALIFerrorbars} shows the absolute atomic oxygen ground state density measured at a distance of 1\,mm distance from the COST jet nozzle as a function of the plasma power. The black dots represent the mean values and the shaded area the standard deviation among the four COST jets. The atomic oxygen density increases under-linear with increasing plasma power from $3\times10^{20}\,\rm{m}^{-3}$ at 0.2\,W to $1\times10^{21}\,\rm{m}^{-3}$ at 1\,W. Our results are in good agreement with previous findings \cite{Willems2019} from molecular mass beam spectrometry and nanosecond TALIF measurements in the early effluent. The observed increase of atomic oxygen density with plasma power in the near effluent is consistent with the measured increase of ozone in the far effluent, see figure \ref{fig:ozonedens}, in view of the efficient chemical conversion of O and O\textsubscript{2} into ozone with increasing reaction time/distance from the nozzle. The standard deviation between the four COST jets for the atomic oxygen density at the nozzle increases with increasing plasma power, e.g. smaller than 5\,\% below 0.5\,W, while increasing up to 13\,\% at 1\,W.  

It should be noted that the stated absolute atomic oxygen densities are prone to an additional systematic uncertainty of more than 20\,\% due to the uncertainty of the two-photon excitation cross section ratio that is required for the TALIF calibration \cite{Dobele2005}. This \textit{absolute} error is not shown in figure \ref{fig:TALIFerrorbars}, since it does not affect the \textit{relative} comparison of the investigated COST jets as presented here.

\subsection{Bactericidal Activity}

Surface decontamination is one application of atmospheric pressure low temperature plasmas (LTPs). Due to their low temperatures, thermally sensitive objects, including skin and other biological tissues can be treated with LTPs to remove bacterial burdens \cite{Weltmann2017,Alkawareek2012,Alkawareek2012b}. As the development of the COST reference microplasma jet is intended to aid the advancement of the field of plasma medicine, it is important to understand how similarly the jets perform to one another, in a biological assay. To do this, a basic bacterial killing assay was used, to determine (a) the efficacy of bacterial killing by the COST jet and (b) how similar the killing ability is between different COST jets. For this, the non-pathogenic \textit{E. coli} MG1655 strain was plated onto LB agar plates and subjected to two minute treatment with the COST jets, using the protocol outlined above. The treatment effects were quantified in two ways. 
First, the area of inhibition (AOI) was measured, and secondly, surviving colonies were counted to calculate the log reduction of bacteria in the AOI. Experiments were repeated in triplicate for each jet, and the mean and standard deviation calculated.

The experimental process was to allow the jet to warm up for 30\,minutes, carry out treatments, then repeat the process for another jet. As this was a fairly lengthy process, treatments with the first jet were then repeated to check that the bacteria had not changed over time due to being left in culture for longer on the bench. These checks showed that there was no difference in the AOI or log reduction in bacteria between the first time the jet was used for treatments, and the second time approximately an hour later (data not shown). The chosen treatment time was 2\,minutes as this was long enough to give a consistent sized AOI, without making the treatment process too long.

\begin{figure}[ht!]
\centering
\includegraphics[width=\linewidth]{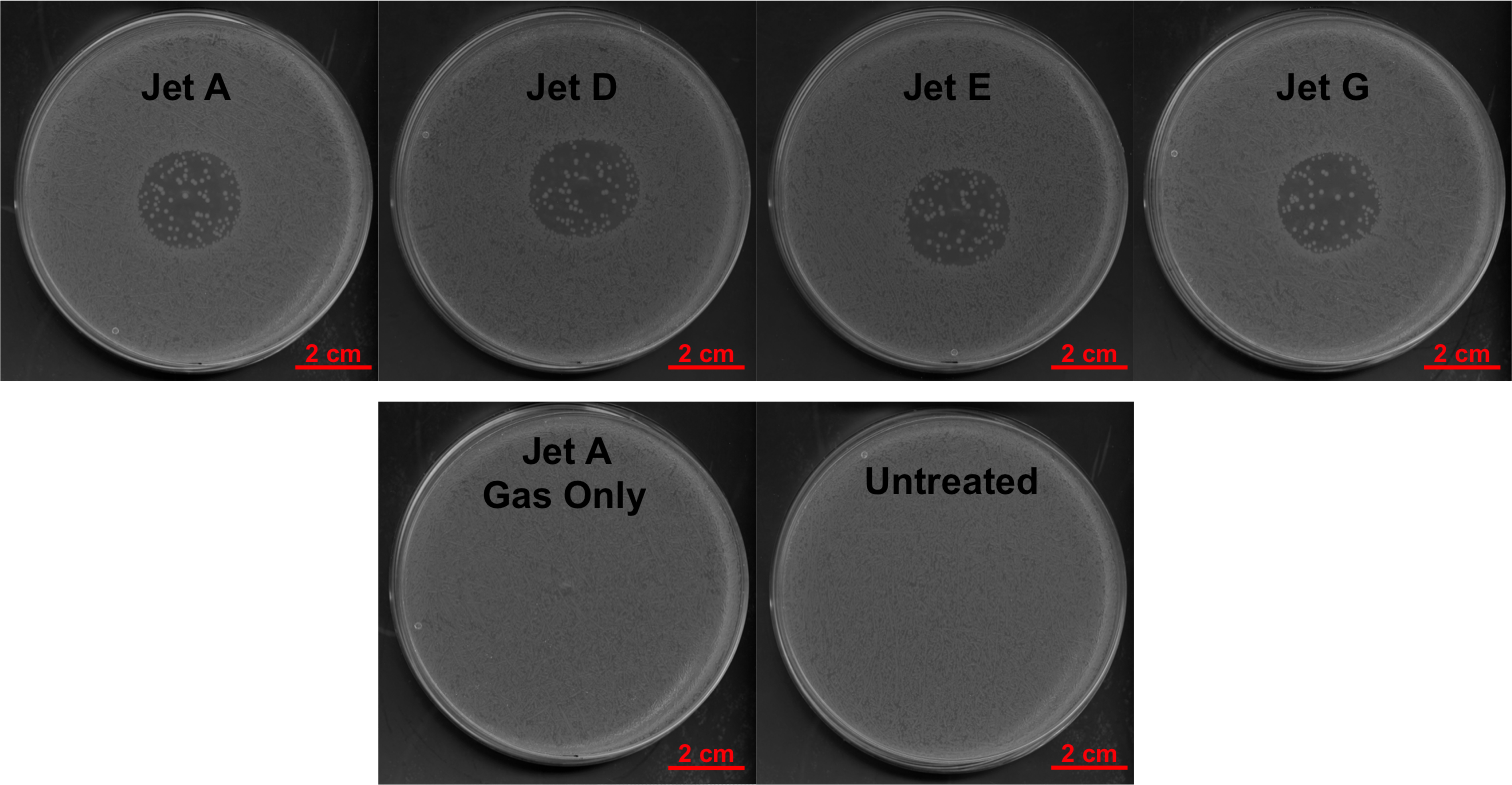}
\caption{Representative images of bacterial plates treated by COST jets.
100 $\mu$L of \textit{E. coli} MG1655 at approximately 8 x 10\textsuperscript{6} cfu/mL were plated onto LB agar plates and exposed to a 2\,minute treatment by COST jet. The plasma power was kept at 0.3\,W, and the feed gas was 1\,slm helium with 0.5\,\% oxygen admixture.
The top panel shows a representative plate for each jet, plates following 2\,minute COST jet treatment and overnight incubation. Representative control plates are shown in the bottom panel.
The gas-only control was also treated for 2\,minutes but the plasma power was turned off (therefore only the helium/oxygen gas was incident on the sample), and the untreated control was plated identically to all the other treatment plates, but did not receive a COST jet treatment.
For each jet, treatments were carried out in triplicate.}
\label{fig:RepImages}
\end{figure}

Representative images of treated and control plates are shown in figure \ref{fig:RepImages}. Treatment with each of the four jets gives circular AOI which appear similar across all the jets. The position of the AOI differs between jets as a result of the each plate not being placed exactly centrally below the plasma nozzle each time. As well as showing similar AOI across all jets, the number of survivor colonies across all of the jets also appears similar in figure \ref{fig:RepImages}. Surviving colonies are thought to occur due to imperfections in the plated bacterial monolayer resulting in some cells being in multiple layers. As a result, cells in upper layers could shield bacteria in lower layers, and prevent their killing by plasma treatment. A gas-only treatment control was also included to confirm the AOI seen in plasma-treated plates was due to plasma, not just the gas flow. The representative image of the gas-only control in the lower left image of figure \ref{fig:RepImages} confirms that the gas flow does not induce an AOI. As expected, the untreated control in the lower right image of figure \ref{fig:RepImages} shows an even coverage of bacteria, with no clear regions. As well as acting as an experimental control, the untreated control plates were also checked to confirm that the bacterial plating methods were good.

\begin{figure}[ht!]
\centering
\includegraphics[width=0.539\linewidth]{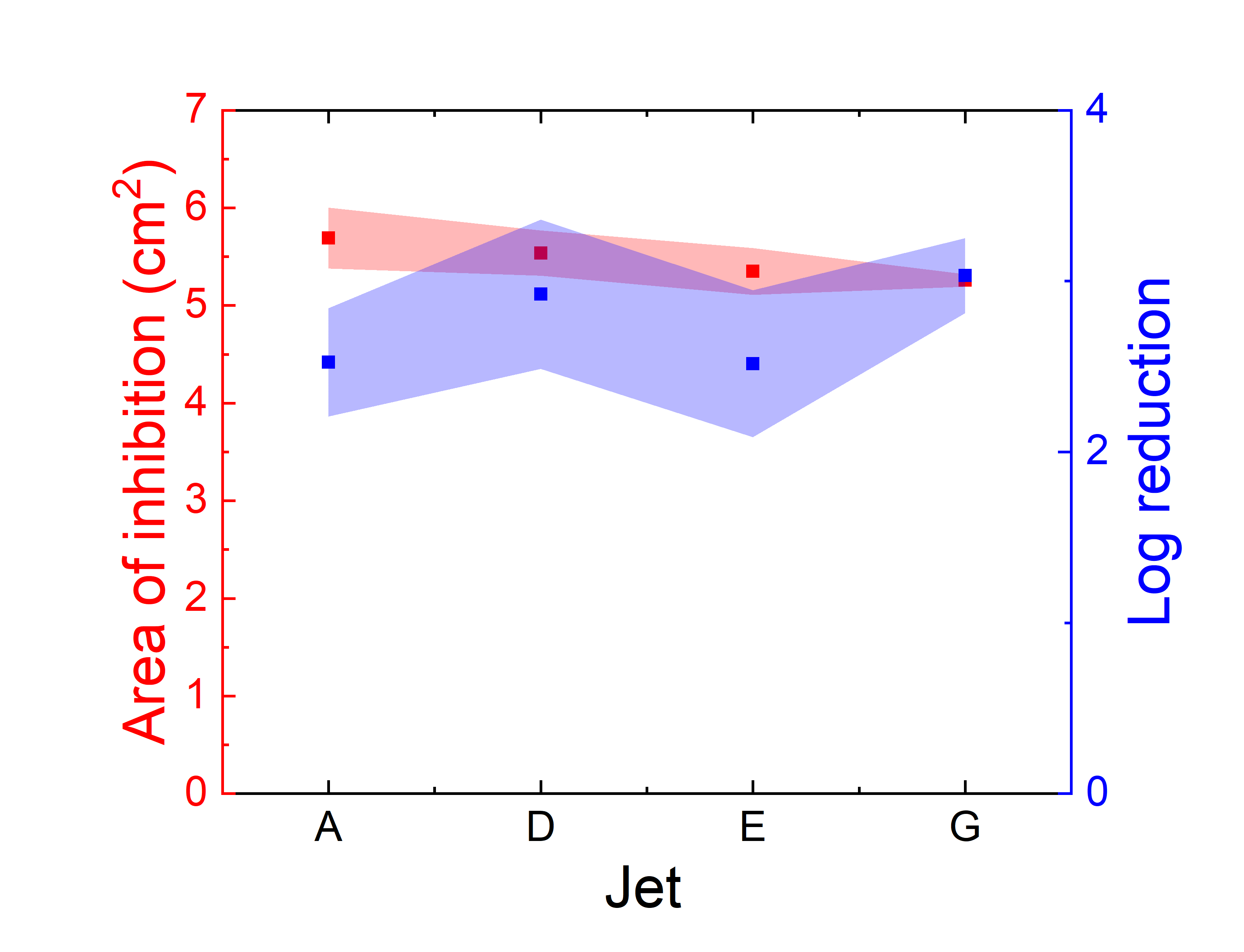}
\caption{Figure showing area of inhibition (AOI) and bacterial log reduction following treatment with the different COST jets. The red points show the average AOI induced by each jet, with the error bars showing the standard deviation. The blue points show the mean log reduction in \textit{E. coli} MG1655 colony forming units (cfu) following treatment with each jet, with the error bars showing standard deviation. For each jet, treatments were carried out in triplicate. Conditions were as stated in figure \ref{fig:RepImages}.}
\label{fig:biostuff}
\end{figure}

To quantify the comparability between jets in the bacterial killing assay, the AOIs and number of survivor colonies were compared. The mean and standard deviation of the AOIs across all four jets, following three repeats for each jet is shown in figure \ref{fig:biostuff} by the red points, axis and shaded area. The AOIs for each of the jets were very similar, with means ranging from 5.3 - 5.7\,cm\textsuperscript{2}. The log reduction of bacteria in the AOIs for each jet were also calculated, and is shown in figure \ref{fig:biostuff} by the blue data points, right axis and blue shaded area. Similar to the AOIs, the log reduction in bacteria due to treatment with each jet also appears to be consistent across all the jets, showing approximately 2.5\,-\,3\,log reduction by each jet. There is some variation between jets, however, this variation is generally smaller than the variation seen within each jet. In general, the four jets have similar abilities to kill our \textit{E. coli} model bacteria. 

The biological effects induced by LTP treatment are expected to be due to the synergies between the relatively high fluxes of reactive oxygen and nitrogen species, and UV delivered to the biological target, as investigated in detail by Schneider et al. and Lackmann et al. \cite{Schneider2011,Schneider2011a,Lackmann2013} and discussed above. The treatments were carried out in a perspex box to reduce any effects of air flow in the laboratory interfering with the RONS delivery to the treated bacteria. To prevent excessive build up of long-lived, toxic species, such as $NO_x$ and $O_3$, an extractor fan was attached to the box, and the $NO_x$ and $O_3$ concentrations in the box were monitored using commercially available monitors (2B Technologies: Model 106-L $O_3$ and Model 405 nm $NO_2/NO/NO_x$). This monitoring showed that these species did not increase appreciably over the treatment time, suggesting that the bacterial killing effects were due to the actual plasma treatment rather than any build up of species in the box. This is further confirmed by the definition of the AOI, which suggests local effects are due to the direct plasma treatment, rather than as a result of the box environment.

\section{Summary and conclusions}

In this study, we compared four COST jet devices constructed to the same nominal specifications, with regard to their actual performance in terms of the internal dissipated plasma power and the resulting external quantities, like effluent gas temperature, sample surface temperature, optical plasma emission, ozone output in the far effluent, generated atomic oxygen density in the near effluent, as well as their bactericidal activity. 

The standard deviations found, for each measured external quantity among the four COST jets, respectively, were below the stated standard deviation of 15\,\% for the internal plasma power. The uncertainty of the power measurement itself, rather than actual differences between the individual COST jets, contributes the greatest to this 15\.\% value, see \cite{Golda2019}.

The effluent gas temperature and the sample surface temperature are critical parameters for the treatment of heat-sensitive material such as biological tissue. Both were found to agree well within a narrow standard deviation of about 3\,\%. It was found that restricting the plasma power to 0.3\,W limits the temperature to the critical value of $37\,^\circ\rm{C}$. In that sense the COST jet can be used safely for the treatment of biological tissue without the necessity of monitoring the temperature of gas effluent or sample surface.

Absolute densities of reactive oxygen species, known to play a key-role in many surface and biological sample treatment processes, were measured. At the reduced plasma power of 0.3\,W, an atomic oxygen density of about $4\times10^{20}$\,\rm{m}$^{-3}$ at 1\,mm distance from the jet nozzle and an ozone density of $1.3\times10^{21}$\,\rm{m}$^{-3}$ in the far effluent were found. The contributions of measurement accuracy and difference between the four COST jets to the observed overall standard deviations, about 10\,\% for the atomic oxygen density and about 3\,\% for the ozone density, is about equal.

A performance study of the four COST jets using the bactericidal assay was conducted. It was found that the achieved bacterial log reduction differed less between the individual COST jets, than between different experiments with one COST jet.

In conclusion, the COST Reference Microplasma Jet is a simple, inexpensive and robust plasma source. Results obtained with four exemplar devices show consistently less than 15\,\% differences, when the internal plasma power is used as the control parameter. This makes the COST jet design a suitable candidate to act as a reference source for scientists working in this field to compare their results as directly as possible.

\section{Acknowledgements}

The authors would like to thank other members of the COST\,MP1101 consortium (P.\,Beijer, A.\,Sobota, G.\,Kroesen, N.\,St.\,J.\,Braithwaite, S.\,Reuter, and M.\,M.\,Turner) for fruitful discussions and for providing their COST jet devices. Richard Armitage is gratefully acknowledged for his technical help.

Part of this work was funded by the DFG within PAK\,816  \emph{PlaCiD: Plasma-cell interactions in Dermatology} and CRC\,1316 \emph{Transient Atmospheric Plasmas: from plasmas to liquids to solids}. We would also like to acknowledge funding from the UK\,EPSRC (EP/K018388/1 \& EP/H003797/2) and the Wellcome Trust 4-year PhD programme [WT095024MA]: \emph{"Combating Infectious Disease: Computational Approaches in Translational Science"}.

\section{References}

\bibliography{COSTjetpaper}
\bibliographystyle{iopart-num}

\end{document}